\documentclass[twocolumn,prl,showpacs,preprintnumbers,amsmath,amssymb]{revtex4}
\usepackage{graphicx}% Include figure files
\usepackage{dcolumn}% Align table columns on decimal point
\usepackage{bm}% bold math
\begin{document}
\title{Berry's phase at quantum vacuum level}

\author{Jian Qi Shen $^{1,2}$}

\email{jqshen@coer.zju.edu.cn}

\affiliation{$^{1}$  Center for Optical and Electromagnetic
Research, Joint Research Center of Photonics of the Royal
Institute of Technology (Sweden) and Zhejiang University, Zhejiang
University, Hangzhou Yuquan 310027, People's Republic of China\\
$^{2}$ Zhejiang Institute of Modern Physics and Department of
Physics, Zhejiang University, Hangzhou 310027, People's Republic
of China}

\date{\today}

\begin{abstract}
The quantum vacuum contribution to Berry's geometric phase of
photon fields inside a noncoplanarly curved (coiled) fiber is
considered by means of the second-quantization formulation. It is
shown that the quantum vacuum Berry's phases of left- and
right-handed circularly polarized light have the equal magnitudes
but opposite signs, and are therefore eliminated entirely by each
other. In order to realize such a novel vacuum effect, a scheme to
separate the quantum vacuum Berry's phase of one polarized light
from another by using the chiral medium fiber is suggested. We
think the present study might be the first treatment for the time
evolution of vacuum zero-point energy.
\end{abstract}
%keywords{Suggested keywords}%Use showkeys class option if keyword
                              %display desired

\maketitle

During the past two decades, the topics on geometric phases and
time-dependent quantum systems \cite{Berry} have attracted
extensive attention of a large number of investigators in various
fields, including quantum optics \cite{Gong}, condensed matter
physics \cite{Taguchi}, nuclear physics \cite{Wagh}, gravity
theory \cite{Furtado} as well as molecular physics (molecular
chemical reaction) \cite{Shenprb}. Recently, geometric phase has
practical applications in the subjects of quantum decoherence and
geometric (topological) quantum computation \cite{Wu}. As for the
purely theoretical aspect of geometric phases, historically, Berry
established a semiclassical connection between the quantal
geometric phase $\gamma$ \cite{Berry} and the classical Hannay's
angle $\Delta\theta_{l}\left(I\right)$ \cite{Hannay} as follows:
$\Delta\theta_{l}=-\partial \gamma/\partial n_{l}$. Here, the
geometric phase is associated with the eigenstates with quantum
numbers $n=\left\{n_{l}\right\}$, and the Hannay's angle
$\Delta\theta_{l}$ is a shift in the $l$th angle variable for
motion round a phase-space torus with actions
$I=\left\{I_{l}\right\}$ \cite{Berry2}. According to Berry's
relation, one can obtain
$\gamma=-n_{l}\Delta\theta_{l}+\gamma_{0}$ (here the summation
over the repeated indices is implied), where $\gamma_{0}$ is an
integral constant. Thus, a new question that might have never been
considered before is left to us: what is the physical meanings of
the nonvanishing integral constant $\gamma_{0}$ (should such
exist)? We believe that such an integral constant may have close
relation to the quantum vacuum contribution to the time-dependent
quantum system, namely, $\gamma_{0}$ may be viewed as Berry's
phase at quantum vacuum level, which arises from the quantum
fluctuation of vacuum.

Quantum vacuum effects have so far captured intensive attention of
many researchers in quantum optics, quantum field theory and
atomic (molecular) physics \cite{Casimir,Bethe,Spruch}. These
effects are as follows: the Casimir effect \cite{Casimir},
anomalous magnetic moment of electron, vacuum polarization, Lamb's
shift \cite{Bethe} as well as Casimir-Polder potentials
\cite{Spruch}. More recently, some new vacuum effects caused by
quantum fluctuation in the inhomogeneous and anisotropic
electromagnetic materials have been predicted theoretically or
observed experimentally. These include the dramatic modification
of spontaneous emission in photonic crystals \cite{Yablonovitch}
and EIT media (multilevel atomic ensemble) \cite{Zhu},
magnetoelectric birefringences of quantum vacuum \cite{Rikken} and
vacuum contribution to the momentum of anisotropic media ({\it
e.g.}, magnetoelectric materials) \cite{Feigel}. Even though some
researchers investigated the problem of light propagation in a
noncoplanarly curved (coiled) fiber by means of various methods,
including the classical electromagnetics, differential geometry
(parallel transport) as well as the first-quantization formulation
\cite{Tomita,Chiao}, to the best of our knowledge, the treatment
based on the second-quantization formulation has so far never been
considered in references. In this Letter, we will study the wave
propagation of the coiled light (and hence the quantum vacuum
contribution to Berry's phase) inside the framework of second
quantization, namely, a physical realization will be proposed for
photon Berry's phase at vacuum level, which originates from the
quantum vacuum fluctuation. It should be noted that since the
left- and right-handed (LRH) circularly polarized photons acquire
the quantum-vacuum Berry's phases with equal magnitudes but
opposite signs, such vacuum contributions to the LRH polarized
light are always eliminated completely inside the {\it isotropic}
media and may therefore have no observable effects experimentally.
However, in some certain {\it anisotropic} media such as chiral
media, gyrotropic (gyroelectric or gyromagnetic) media and
magnetoelectric materials, the quantum vacuum contribution may no
longer be exactly cancelled for the LRH circularly polarized light
since the noncompensation effect \cite{Feigel} of a pair of LRH
vacuum modes will arises in these {\it anisotropic} media. We
suggest an experimentally feasible scheme to test this new vacuum
effect by using the noncoplanar (coiled) chiral-medium fiber in
Tomita-Chiao experiment \cite{Tomita}.

First, we consider the wave propagation inside the noncoplanar
fiber. According to the Maxwellian equations, the field vectors
${\bf G}_{\pm}={\bf E}\pm iv{\bf B}$ agree with the following
equations
\begin{equation}
\nabla\times{\bf G}_{\pm}=\pm\frac{i}{v}\frac{\partial}{\partial
t}{\bf G}_{\pm}     \label{Maxwellian}
\end{equation}
with the phase velocity $v=c/n$, where $n$ denotes the optical
refractive index of the linear electromagnetic medium. Consider a
planar electromagnetic wave whose field vector $G_{+}\propto
\exp\left(i{\bf k}\cdot{\bf r}\right)$. Thus, Eq.
(\ref{Maxwellian}) can be rewritten as ${\bf k}\times{\bf
G}_{+}=(1/v)(\partial/\partial t){\bf G}_{+}$. By using the
relations ${\bf S}\times{\bf S}=i\hbar{\bf S}$ and $\left[{\bf
a}\cdot{\bf S}, {\bf b}\cdot{\bf S}\right]=i\hbar \left({\bf
a}\times{\bf b}\right)\cdot{\bf S}$, where ${\bf S}$ denotes the
spin operator of the photon field, one can further obtain
$\left[v{\bf k}\cdot{\bf S}, {\bf G}_{+}\cdot{\bf S}\right]=i\hbar
(\partial/\partial t){\bf G}_{+}\cdot{\bf S}$. Further analysis
shows that the field vectors ${\bf G}_{\pm}$ (or the operator
${\bf G}_{\pm}\cdot{\bf S}$) satisfy (for simplicity, the
subscript ${\pm}$ will be omitted)
\begin{equation}
\frac{\partial}{\partial t}{\bf G}\cdot{\bf
S}+\frac{1}{i\hbar}\left[{\bf G}\cdot{\bf S}, v{\bf k}\cdot{\bf
S}\right]=0,     \label{Liouville}
\end{equation}
which is just a form of Liouville-von Neumann equation $
\frac{\partial}{\partial t}{I}+({i\hbar})^{-1}\left[I, H\right]=0$
\cite{Riesenfeld}. Here, $I$ and $H$ denote the Lewis-Riesenfeld
invariant \cite{Riesenfeld} and the Hamiltonian of the quantum
system, respectively. It follows from Eq. (\ref{Liouville}) and
the Liouville-von Neumann equation that the operator $v{\bf
k}\cdot{\bf S}$ can be regarded as the effective interaction
Hamiltonian that characterizes the wave propagation of circularly
polarized light inside the noncoplanar fiber. Here, ${\bf k}$
denotes the wave vector of one-dimensional longitudinal
propagation along the guiding direction. As the relation between
the frequency $\omega$ and the modulus of wave vector of the
electromagnetic wave in a linear medium is $\omega=kv$, the
effective Hamiltonian $v{\bf k}\cdot{\bf S}$ can be rewritten as
$\omega {\bf n}\cdot{\bf S}$, where the unit vector ${\bf n}\equiv
{\bf k}/k$, which can be expressed as $\left(\sin\theta
\cos\varphi, \sin\theta\sin\varphi, \cos\theta\right)$. In the
noncoplanar fiber, the angle displacements, $\theta
\left(t\right)$ and $\varphi \left(t\right)$, in the
three-dimensional spherical polar coordinate system are the
time-dependent functions. Note that here $\theta \left(t\right)$
and $\varphi \left(t\right)$ can characterize the geometric shape
of the noncoplanarly curved fiber, since it was assumed that in
waveguides (fibers) for photons (or their atomic counterparts)
\cite{Stenholm}, the wave packet describing the propagation
follows the channel smoothly and without too much distortion. Such
assumptions derive from some underlying smoothness of the guiding
structures and thus the adiabaticity will hold in the propagation
process \cite{Stenholm}.

In view of the above discussion, the eigenvalue equation of
instantaneous Hamiltonian $H(t)=\omega {\bf n}(t)\cdot{\bf S}$
governing the wave propagation of the light in the noncoplanarly
curved fiber reads $ \omega {\bf n}(t)\cdot{\bf S}|{\bf k}(t), \pm
\rangle=E_{\pm}|{\bf k}(t), \pm \rangle$. If the initial condition
for the photon wave vector is ${\bf k}(t=0)=\left(0,0,k\right)$
(and hence $\theta(t=0)=0$), then the solutions of this equation
are of the form $ |{\bf k}(t), \pm \rangle=V(t)|\pm \rangle$,
where $|\pm\rangle=|{\bf k}(t=0), \pm \rangle$, and the
time-dependent unitary operator in the solutions is
\begin{equation}
V(t)=\exp\left\{\left[-\frac{\theta(t)}{2\hbar}e^{-i\varphi(t)}\right]S_{+}
-\left[-\frac{\theta(t)}{2\hbar}e^{i\varphi(t)}\right]S_{-}\right\}.
\end{equation}
Here the operators $S_{\pm}=S_{1}\pm iS_{2}$. Further calculation
shows that the explicit expressions for the above solutions take
the form
\begin{eqnarray}
& &    |{\bf k}(t), +\rangle=\left[\cos\frac{\theta(t)}{2}|+\rangle+e^{i\varphi(t)}\sin\frac{\theta(t)}{2}|-\rangle\right],                       \nonumber    \\
& &      |{\bf
k}(t),-\rangle=\left[\cos\frac{\theta(t)}{2}|-\rangle-e^{-i\varphi(t)}\sin\frac{\theta(t)}{2}|+\rangle\right].
\end{eqnarray}
Note that for the propagating electromagnetic wave in the
noncoplanar fiber, the cyclic evolution of photon wavefunction
will yield the original state plus a phase shift \cite{Berry},
which is a sum of a dynamical phase $-\gamma_{\pm}^{({\rm d})}(T)$
and a geometric phase shift $\gamma_{\pm}^{({\rm g})}(T)$, namely,
if the initial state is $|\pm \rangle$, then the state at $T$ is
\begin{equation}
|{\bf k}(T), \pm \rangle=\exp\left[i\gamma_{\pm}^{({\rm
g})}(T)\right]\exp\left[-i\gamma_{\pm}^{({\rm
d})}(T)\right]V(T)|\pm \rangle,
\end{equation}
where $\gamma_{\pm}^{({\rm
d})}(T)=\hbar^{-1}\int^{T}_{0}\langle\pm
|V^{\dagger}(t')H(t')V(t')|\pm\rangle {\rm d}t'$ and
$\gamma_{\pm}^{({\rm g})}(T)=\int^{T}_{0}\langle\pm
|V^{\dagger}(t')i\partial V(t')/\partial t'|\pm\rangle {\rm d}t'$.
Here the explicit expressions for the geometric phase shift
$\gamma_{\pm}^{({\rm g})}(T)=-\hbar^{-1}\langle \pm|
S_{3}|\pm\rangle\int^{T}_{0}\dot{\varphi}(t')\left[1-\cos\theta(t')\right]{\rm
d}t'$, where dot denotes the time derivative.

In the following, we assume that the noncoplanarly curved fiber
has the shape of coil \cite{Berry3}. Consider a typical case where
the precessional frequency $\dot{\varphi}$ of photon moving on the
fiber helicoid is constant (denoted by $\Omega $) and the
nutational frequency $\dot{\theta}$ vanishing. After one period
($T=2\pi/\Omega$) that it takes to complete a precessional cycle
in the coiled fiber, the cyclic geometric phase (Berry's phase
shift) is $ \gamma_{\pm}^{({\rm g})}(T)=-{2\pi
\left(1-\cos\theta\right)}\langle \pm| S_{3}|\pm\rangle/{\hbar}$,
where $2\pi(1-\cos\theta)$ is an expression for the solid angle
subtended at the center by a curve traced by the photon wave
vector in the propagation process inside the coiled fiber.

Now we consider the expectation value, $\left\langle \pm \right|
S_{3}\left| \pm \right\rangle$, of the third component of photon
spin operator in $\gamma_{\pm}^{({\rm g})}(T)$. Substitution of
the Fourier expansion series of three-dimensional magnetic vector
potentials ${\bf A}({\bf x},t)$ into the expression
$S_{ij}=-\int{(\dot{A}_{i}A_{j}-\dot{A}_{j}A_{i})}{\rm d}^{3}{\bf
x}$  \cite{Bjorken} for the spin operator of the photon field
yields the monomode spin operator $
S_{3}=({i}\hbar/{2})[a(k,1)a^{\dagger}(k,2)-a^{\dagger}(k,1)a(k,2)
-a(k,2)a^{\dagger}(k,1)+a^{\dagger}(k,2)a(k,1)] $ with
$a^{\dagger}(k,\lambda)$ and $a(k,\lambda)$ ($\lambda=1, 2$) being
the creation and annihilation operators of polarized photons
corresponding to the two mutually perpendicular polarization
vectors. Here, $\lambda$ denotes the label of the polarization
vectors \cite{Bjorken}. In what follows, we will define the
creation and annihilation operators, $a_{R}^{\dagger}(k)$,
$a_{L}^{\dagger}(k)$, $a_{R}(k)$, $a_{L}(k)$, of right- and
left-handed circularly polarized light
 \cite{Bjorken}, which are expressed in terms of
$a^{\dagger}(k,\lambda)$ and $a(k,\lambda)$, {\it i.e.},
$a_{R}^{\dagger}(k)=[a^{\dagger}(k,1)+ia^{\dagger}(k,2)]/\sqrt{2}$,
$a_{R}(k)=[a(k,1)-ia(k,2)]/\sqrt{2}$,
$a_{L}^{\dagger}(k)=[a^{\dagger}(k,1)-ia^{\dagger}(k,2)]/\sqrt{2}$
and $a_{L}(k)=[a(k,1)+ia(k,2)]/\sqrt{2}$. Thus, the spin operator
of monomode photon can be rewritten in terms of the creation and
annihilation operators of left- and right-handed polarized
photons, {\it i.e.}, $
S_{3}=\left[a_{R}(k)a_{R}^{\dagger}(k)+a_{R}^{\dagger}(k)a_{R}(k)\right]\hbar/2
-\left[a_{L}(k)a_{L}^{\dagger}(k)+a_{L}^{\dagger}(k)a_{L}(k)\right]\hbar/2$,
which can also be rewritten as the form $
S_{3}=\left[a_{R}^{\dagger}(k)a_{R}(k)+{1}/{2}\right]\hbar-\left[a_{L}^{\dagger}(k)a_{L}(k)+{1}/{2}\right]\hbar
$. Hence, Berry's phases (at second-quantization level) of left-
and right-handed circularly polarized light accumulated in one
precessional period ($T=2\pi/\Omega$) are $ \gamma^{({\rm
g})}_{n_{\rm
L}}(T)=+2\pi\left(1-\cos\theta\right)\left(n_{L}+{1}/{2}\right)$
and $ \gamma^{({\rm g})}_{n_{\rm
R}}(T)=-2\pi\left(1-\cos\theta\right)\left(n_{R}+{1}/{2}\right)$,
where $n_{\rm L}$ and $n_{\rm R}$ denote the occupation numbers of
left- and right-handed polarized photons, respectively.

It is physically interesting to consider the connection between
the second- and first-quantized Berry's phases. Such a connection
can be demonstrated by taking account of the expectation value of
the operator of electric field strength in the coherent state:
specifically, the coherent state of the polarized light (say, the
right-handed polarized light) at $t=T$ can be constructed in terms
of the photon states \cite{Gao} $\exp\left[i\gamma_{n_{\rm
R}}(T)\right]|n_{\rm R}, T\rangle$, {\it i.e.},
\begin{equation}
|\alpha,
T\rangle=\exp\left(-\frac{\alpha^{\ast}\alpha}{2}\right)\sum^{\infty}_{n_{\rm
R}=0}\frac{\alpha^{n_{\rm R}}}{\sqrt{n_{\rm
R}!}}\exp\left[i\gamma_{n_{\rm R}}(T)\right]|n_{\rm R}, T\rangle,
\label{coherent}
\end{equation}
where $\gamma_{n_{\rm R}}(T)=\gamma^{({\rm g})}_{n_{\rm
R}}(T)-\gamma^{({\rm d})}_{n_{\rm R}}(T)$ and $|n_{\rm R},
T\rangle=V(T)|n_{\rm R}\rangle$. The expectation value of the
operator $\hat{q}=\left(a_{{\rm R}}+a^{\dagger}_{{\rm
R}}\right)/\sqrt{2}$ of the electric field strength in the
coherent state (\ref{coherent}) is
\begin{equation}
\langle\hat{q}\rangle
(T)=\frac{1}{\sqrt{2}}\left\{\alpha^{\ast}\exp\left[-i\Delta_{\rm
R}(T)\right]+\alpha\exp\left[i\Delta_{\rm R}(T)\right]\right\},
\label{expectation}
\end{equation}
where $\Delta_{\rm R}(T)=\gamma_{n_{\rm R}+1}(T)-\gamma_{n_{\rm
R}}(T)$, which is a phase shift independent of the photon
occupation number $n_{\rm R}$. Thus, the geometric phase
contribution in the phase shift $\Delta_{\rm R}(T)$ is
\begin{equation}
\gamma^{({\rm g})}_{\rm R}=-2\pi\left(1-\cos\theta\right),
\label{result}
\end{equation}
which is a first-quantized Berry's phase acquired by the
electromagnetic wave in the cyclic process. It should be noted
that the result (\ref{result}) obtained here is consistent with
the previous studies based on the differential geometry method,
Maxwellian equations and quantum mechanics (first quantization)
\cite{Tomita,Chiao}. So, we think the treatment presented in this
Letter is self-consistent. Apparently, there exists no Berry's
phases at quantum vacuum level in the expression (\ref{result})
since $\Delta_{\rm R}(T)$ in (\ref{expectation}) is merely a phase
shift at first-quantization level.

It follows from the expressions for $\gamma^{({\rm g})}_{n_{\rm
L}}(T)$ and $\gamma^{({\rm g})}_{n_{\rm R}}(T)$ that Berry's
phases of the left- and right-handed polarized vacuum modes are
always having the equal magnitudes but opposite signs, {\it i.e.},
$\gamma^{({\rm g})}_{0_{\rm L}}(T)=+\pi\left(1-\cos\theta\right)$
and $\gamma^{({\rm g})}_{0_{\rm
R}}(T)=-\pi\left(1-\cos\theta\right)$. Unfortunately, such a novel
vacuum effect may not manifest the observable phenomena
experimentally. Indeed, historically, Berry's phases at quantum
vacuum level inside the curved fiber have not been detected in the
previous experiments \cite{Tomita} just due to such a
cancellation.

However, it is reasonably believed that since the vacuum mode
structures can be influenced dramatically by the {\it anisotropic}
\cite{Feigel} (and {\it inhomogeneous} \cite{Yablonovitch})
environment, some quantum vacuum effects will unavoidably arise in
anisotropic (and inhomogeneous) media. Here, we will take into
account the wave propagation in the chiral medium fiber, in which
the quantum vacuum contribution to the left- and right-handed
polarized fields will no longer be eliminated by each other. A
homogeneous chiral medium can be characterized by the following
constitutive relations \cite{Cheng}
\begin{equation}
{\bf D}=\epsilon\epsilon_{0}{\bf E}+i\zeta {\bf B}, \quad {\bf
H}=i\zeta{\bf E}+\frac{{\bf B}}{\mu_{0}},    \label{constitutive}
\end{equation}
where $\epsilon$ and $\zeta$ denote the relative permittivity and
the chirality parameter, respectively. As a preliminary
consideration, let us first analyze the problem of field
quantization in chiral media. By expanding the three-dimensional
electromagnetic vector potentials ${\bf A}({\bf x},t)$ as a
Fourier series ${\bf A}({\bf x},t)=\sum_{l}{\bf A}_{l}({\bf
x})f_{l}(t)$, where ${\bf A}_{l}({\bf x})={\bf
u}_{l}\exp\left(i{\bf k}_{l}\cdot{\bf x}\right)$ (${\bf u}_{l}$ is
a unit vector) and
$f_{l}(t)=|f_{l}|\exp\left(-i\omega_{l}t\right)$, one can obtain
the expression $
W=2\left(\epsilon\epsilon_{0}+\mu_{0}\zeta^{2}\right)V\sum_{l}\omega^{2}_{l}f_{l}^{\ast}f_{l}
$ for the energy of electromagnetic field in the chiral medium,
where $V$ denotes the volume of the medium. It follows that for
the electromagnetic energy density in the chiral medium, the role
of the chirality parameter $\zeta$ is just to make a correction to
the scalar factor $\epsilon\epsilon_{0}$. Thus, we believe that
the conventional mathematical formalism of the field quantization
in isotropic homogeneous electromagnetic media can apply in chiral
media: specifically, by introducing
$q_{l}=\left[\left(\epsilon\epsilon_{0}+\mu_{0}\zeta^{2}\right)V\right]^{1/2}\left(f_{l}+f_{l}^{\ast}\right)$
and
$p_{l}=-i\left[\left(\epsilon\epsilon_{0}+\mu_{0}\zeta^{2}\right)V\right]^{1/2}\omega_{l}\left(f_{l}-f_{l}^{\ast}\right)$,
the above expression for the field energy in the chiral medium can
be rewritten as
$W=\sum_{l}\left(p^{2}_{l}+\omega^{2}_{l}q^{2}_{l}\right)/2$. By
using the quantization technique, in which the annihilation and
creation operators of photon fields are defined as follows:
$b_{l}=(2\hbar\omega_{l})^{-1/2}\left(\omega_{l}\hat{q}_{l}+i\hat{p}_{l}\right)$,
$b_{l}^{\dagger}=(2\hbar\omega_{l})^{-1/2}\left(\omega_{l}\hat{q}_{l}-i\hat{p}_{l}\right)$,
the formula of the field energy in the chiral medium can be
written in the form
$W=(1/2)\sum_{l}\hbar\omega_{l}\left(b_{l}^{\dagger}b_{l}+b_{l}b_{l}^{\dagger}\right)$.
This, therefore, means that the treatment and results of the
quantum vacuum contribution to Berry's phase presented above is
still valid for the case of chiral media.

According to the constitutive relations (\ref{constitutive}), a
time harmonic electromagnetic wave agrees with the wave equation
\begin{equation}
\left(\nabla^{2}-\epsilon\epsilon_{0}\mu_{0}\frac{\partial^{2}}{\partial
t^{2}}\right)\left(E_{1}\pm iE_{2}\right)\mp 2\zeta\mu_{0}k\omega
\left(E_{1}\pm iE_{2}\right)=0.
\end{equation}
Therefore, the wave vectors $k_{\rm L,R}$ corresponding to the
wave amplitudes $E_{1}\pm iE_{2}$ fulfill the two quadratic
equations $
  k^{2}_{\rm R}+2\zeta \mu_{0}\omega k_{\rm
  R}-\epsilon\epsilon_{0}\mu_{0}\omega^{2}=0 $ and $
   k^{2}_{\rm L}-2\zeta \mu_{0}\omega k_{\rm
L}-\epsilon\epsilon_{0}\mu_{0}\omega^{2}=0$, respectively. The
solutions of the quadratic equations are given by $ k_{\rm
R}=\mu_{0}\omega\left(-\zeta+\sqrt{\zeta^{2}+{\epsilon\epsilon_{0}}/{\mu_{0}}}\right)$
and $ k_{\rm
L}=\mu_{0}\omega\left(+\zeta+\sqrt{\zeta^{2}+{\epsilon\epsilon_{0}}/{\mu_{0}}}\right)$.
Note that here both the negative roots of $k_{\rm L,R}$ have been
ruled out since it is assumed that the wave vectors of the left-
and right-handed polarized light have the same direction in the
propagation process inside the coiled fiber. It is thus clear that
the wave vector $k_{\rm R}$ of the right-handed polarized light is
different from $k_{\rm L}$ of the left-handed polarized light
because of the chirality parameter $\zeta$, and that the
difference between them is $\Delta k=-2\zeta \mu_{0}\omega$. This
may lead to the possibility to differentiate between
quantum-vacuum Berry's phases of left- and right-handed polarized
fields, which can be interpreted as follows: the precessional
frequency of electromagnetic wave propagating on the helicoid
reads $\Omega ={2\pi v}/{\sqrt{d^{2}+(4\pi a)^{2}}}$, where $v$
denotes the phase velocity of light inside the coiled fiber; $a$
is the radius of the helix and $d$ the helical pitch length. If
the parameter $\zeta$ in the constitutive relations
(\ref{constitutive}) is a small number (compared with
$\sqrt{\epsilon\epsilon_{0}/\mu_{0}}$), then the difference
between the precessional frequencies $\Omega_{\rm R}$ and
$\Omega_{\rm L}$ is $ \Delta\Omega=4\zeta{\pi
c^{2}\mu_{0}}/[{\epsilon\sqrt{d^{2}+(4\pi a)^{2}}}]$. Meanwhile,
the difference between the precessional periods of right- and
left-handed circularly polarized light propagating on the helicoid
of the coiled fiber is $ \Delta T=-2\zeta\mu_{0}\sqrt{d^{2}+(4\pi
a)^{2}}$. It follows that because of the existence of the
chirality parameter $\zeta$, the precessional frequencies (and
hence the precessional periods) are different between the left-
and right-handed polarized light inside the coiled fiber. This may
enable us to separate one of the circularly polarized light from
another, and thus no longer allows the quantum vacuum contribution
to Berry's phases to be cancelled totally by each other in one
precessional cycle. Hence, such a novel vacuum effect is possible
to be observed in the coiled fiber experiments \cite{Tomita}, if
one utilizes the fibers fabricated by the chiral media instead of
those made of regular (isotropic) media.

To summarize, we considered a new vacuum effect caused by the time
evolution of the zero-point energy of quantum vacuum fluctuation.
As the noncompensation effect of a pair of counter-propagating (or
-spinning) vacuum modes will arise in {\it anisotropic} media
\cite{Feigel}, the physical effects resulting from quantum vacuum
fluctuation of left- and right-handed polarized modes will no
longer be exactly cancelled by each other. This may lead to an
observable effect of the quantum vacuum contribution to Berry's
phase in time-dependent quantum systems. Since Berry's phase at
quantum vacuum level may have close relation to some fundamental
problems, {\it e.g.}, the time evolution of vacuum background,
field quantization, time-dependent field theory as well as cavity
QED, we hope such an interesting vacuum effect would be realized
experimentally in the near future.

\begin{acknowledgments}
This work was supported in part by both the National Natural
Science Foundation of China (Project Nos. $90101024$ and
$60378037$) and the National Basic Research Program (973) of China
(Project No. 2004CB719805).
\end{acknowledgments}

\end{document}